\documentclass[a4paper,11pt]{article}
\usepackage{a4}

\usepackage{amsbsy}
\usepackage{amssymb}
\usepackage{amsmath}
\usepackage{epsfig}
\usepackage{fullpage}

\def\dd{\mbox{d}}

\def\O{\Omega}

\def\bra{\langle}
\def\ket{\rangle}
\def\a{\alpha}
\def\b{\beta}
\def\d{\delta}

\def\f{\phi}
\def\F{\Phi}

\def\l{\lambda}

\def\L{\Lambda}
\def\m{\mu}

\def\S{\Sigma}

\def\t{\tau}
\def\th{\theta}

\def\pa{\partial}

\newcommand{\ti}[1]{\tilde{#1}}

\newcommand{\sm}[1]{\mbox{\scriptsize #1}}

\renewcommand{\@}[1]{\sqrt{#1}}
\newcommand{\Tr}{{\mbox{Tr}}\,}
\renewcommand{\le}[1]{\label{#1}\end{eqnarray}}
\newcommand{\be}{\begin{equation}}
\newcommand{\ee}{\end{equation}}
\newcommand{\bea}{\begin{eqnarray}}
\newcommand{\eea}{\end{eqnarray}}
\newcommand{\nn}{\nonumber}

\newcommand{\eq}[1]{(\ref{#1})}
\def\nn{\nonumber\\}

\def\ffract#1#2{\raise .35 em\hbox{$\scriptstyle#1$}\kern-.25em/
\kern-.2em\lower .22 em \hbox{$\scriptstyle#2$}}

\begin{document}

\begin{flushright}
AEI-2003-108\\
UCLA/03/TEP/41\\
ITFA-2004-09\\
{\tt hep-th/0403035}
\end{flushright}
\vskip0.1truecm
\begin{center}
\vskip 2truecm
{\Large\bf The Off-shell M5-brane and Non-Perturbative Gauge Theory}
\vskip 1.5truecm
{\large\bf Jan de Boer${}^{\star}${} and Sebastian de Haro${}^{\dagger}$}\\
\vskip .5truecm
${}^{\star}${\it Institute for Theoretical Physics\\
University of Amsterdam\\
Valckenierstraat 65, 1018 XE Amsterdam, The Netherlands}\\
{\tt jdeboer@science.uva.nl}
\vskip 4truemm
${}^{\dagger}$
{\it Max-Planck-Insitut f\"ur Gravitationsphysik\\
Albert-Einstein-Institut\\
14476 Golm, Germany}\\
{\tt sdh@aei.mpg.de}\\
\vskip 2truemm
\end{center}
\vskip 1truecm
\begin{center}
{\bf \large Abstract}
\end{center}

M5-branes wrapping a holomorphic curve in a Calabi-Yau manifold can be used 
to construct
four-dimensional ${\cal N}=1$ gauge theories. In this paper we will consider 
M5-brane
configurations corresponding to ${\cal N}=2$ theories broken to ${\cal N}=1$ 
by a
superpotential for the adjoint scalar field. These M5-brane configurations 
can
be obtained
by lifting suitable intersecting brane configurations in type IIA, or 
equivalently by
T-dualizing IIB configurations with branes and/or fluxes. We will show that 
turning
on non-trivial expectation values for the glueball superfields corresponds 
to
non-holomorphic deformations of the M5-brane. We compute the superpotential 
and show
it agrees with that computed by Dijkgraaf and Vafa. Several aspects of the 
gauge theory,
such as the appearance of non-holomorphic one-forms with integer periods on 
the
Seiberg-Witten curve, have a natural interpretation from the M5-brane point 
of view. We also explain the interpretation of the superpotential in terms 
of the
twisted (2,0) theory living on the fivebrane.

\newpage

\section{Introduction}

Recently, major progress has been achieved in understanding the
interconnections between F-terms in 4d ${\cal N}=1$ SYM theory, topological
string theory, matrix models, and integrable systems, starting with
\cite{dv1}. In particular, a remarkably simple prescription to determine
the non-perturbative superpotential of certain ${\cal N}=1$ gauge theories
in terms of a matrix integral was given in \cite{dv1}. The superpotential
is expressed as a function of gluino condensate superfields $S_i$, and it
is intriguing that one is in some sense doing off-shell computations, i.e.
away from the supersymmetric minima of the superpotential.

One aspect of this off-shell construction that has been less well
understood is the need to include non-holomorphic quantities when
one goes off-shell. In the type IIB construction of these gauge
theories \cite{civ} one employs a Calabi-Yau manifold with
three-form flux, and the superpotential is of the form $W=\int
H\wedge \Omega$, with $H=H_R + \tau H_{NS}$ the three-form flux
and $\Omega$ the holomorphic three-form. Superficially, the entire
superpotential comes from the anti-holomorphic part of the flux.
This parallels the discussion in field theory \cite{csw1,csw2},
where it was shown that the off-shell variation of the
superpotential involves a closed one-form $T$ that is not
holomorphic. This is because once one has fixed the A- and
B-periods of a holomorphic differential, one cannot independently
turn on expectation values for the gaugino condensates -- as this
corresponds to varying the moduli of the underlying Riemann
surface. If one wants to do the latter, one needs to consider a
non-holomorphic one-form $T$.

In the present paper we wish to study the non-perturbative physics and in 
particular the
superpotential of deformed ${\cal N}=2$ gauge theories
from the M-theory perspective. The description of ${\cal N}=2$ theories in 
terms of an
M5-brane wrapped on the Seiberg-Witten curve was first found in \cite{wi1}, 
by lifting a
suitable intersecting brane configuration in type IIA to M-theory. This was 
generalized
to mass-deformed ${\cal N}=2$ and pure ${\cal N}=1$ theories in 
\cite{hoo,wi2}. For a review
and applications of intersecting brane configurations to gauge dynamics see 
e.g. \cite{brev}.
The M5-brane configuration for ${\cal N}=2$ theories deformed by a general 
superpotential
for the adjoint scalar field was found in \cite{bo}, again by lifting a 
suitable brane
configuration from type IIA to M-theory. The process of lifting a brane 
configuration
from type IIA to M-theory is T-dual to the large $N$ transition of 
\cite{va1}: the type IIA
configuration with intersecting branes is dual to a resolved type IIB 
geometry with
D5 branes wrapping ${\mathbb P}^1$'s, whereas the M-theory configuration is 
dual to the deformed
type IIB geometry with fluxes (see e.g. \cite{tatar}). The field theory 
living on the
world-volume of the M5-brane is not quite that of an ordinary supersymmetric 
four-dimensional
gauge theory, due to the existence of additional light degrees of freedom 
coming from the
KK-modes of the M-theory circle. For this reason the field theory is 
sometimes called MQCD
instead of QCD. However, as long as we consider BPS-like quantities such as 
the F-terms in
the low-energy effective field theory the M5-brane yields the same answer as 
the gauge theory.
The agreement breaks down for non-BPS quantities such as the K\"ahler 
potential of
${\cal N}=1$ theories and higher derivative terms in the ${\cal N}=2$ theory 
\cite{bhoo}.

Since the superpotential is a holomorphic quantity one should be
able to compute it directly from the M5-brane configuration. In
\cite{wi1} an expression for the effective superpotential was
proposed and evaluated in a few examples. This expression is
rather cumbersome to work with (some more examples are worked out
in \cite{nos}). Here we will compute this superpotential in more
general cases, and hopefully shed some light on its
interpretation. In particular, the M-theory setup gives a purely
geometric interpretation of all the quantities involved in the
Dijkgraaf-Vafa proposal, and can be seen as a pure geometric dual
of the type IIB setup with fluxes: all information is encoded in a
surface embedded in ${\mathbb R}^5 \times S^1$. We will give a
precise off-shell definition of the superpotential and we will see
that it reproduces the Dijkgraaf-Vafa result for 4d SYM theory. As
it will turn out, going off-shell is quite subtle, as the
embedding of the fivebrane in the Calabi-Yau will no longer be
holomorphic. We will also discuss how the superpotential arises
from the kinetic term for the scalars in the (2,0) theory living
on the world-volume of the fivebrane via deconstruction. This
generalizes the relation between the (2,0) theory and ${\cal N}=2$
SYM \cite{hlw,bhoo} to the case of ${\cal N}=1$, and in particular
to the precise computation of the F-terms.

In the ${\cal N}=2$ case the fivebrane worldvolume is of the form
${\mathbb R}^4\times\S$, where the Riemann surface $\S$ is holomorphically
embedded in ${\mathbb R}^3 \times S^1$. Thus there is a non-trivial
background value for one of the complex transverse scalars. In the
case of ${\cal N}=1$, two transverse scalars, which we call $t$
and $w$, have a non-trivial configuration, and so we will take
$\S$ to be embedded in ${\mathbb R}^5 \times S^1$ with complex
coordinates $t,w,v$. The complex coordinate $t$ takes values in
${\mathbb C}^{\ast}$ and parametrizes the cylinder ${\mathbb R}\times S^1$, 
whereas
$v,w$ take values in the complex plane. On-shell, the embedding is
given by two algebraic equations \cite{bo}:
\bea
t^2-2P_N(v)t+\L_{N=2}^{2N}&=&0\nn
w^2-2W'_m(v)w-\ti f_{m-1}(v)&=&0~.
\eea
where $P_N,W'_m$ and $\tilde{f}_{m-1}$ are polynomials of degrees $N,m,m-1$.
In order for $t$ and $w$ to be meromorphic with a prescribed singularity
structure, the following factorization formulas must also hold:
\bea \label{rs1}
P_N(v)^2-\L^{2N}_{N=2}&=&S^2_{N-n}(v)y(v)\nn
W'_m(v)^2+\ti f_{m-1}(v)&=&H_{m-n}^2(v)y(v)
\eea
for suitable polynomials $S,H$, and $y(v)$ describes a
Riemann surface of genus $n-1$:
\be
y^2=G_n^2(v)+f_{n-1}(v)~.
\ee
By counting parameters we see that (\ref{rs1}) has the same number
of free parameters as there are equations for fixed $W'_m$. Thus
given a superpotential $W$, these equations will generically have
a discrete set of solutions for all other coefficients. This
factorization agrees with the results from field theory
\cite{bo,civ}. We propose the following relation of the embedding
coordinates $t$ and $w$ to the gauge theory quantities:
\bea \label{per1}
N_i=\frac{1}{2\pi i} \oint_{A_i}{\frac{\dd t}{t}}&& \t+\t_i=
\frac{1}{2\pi i} \int_{\hat B_i}{\frac{\dd t}{t}}\nn S_i=\frac{1}{2\pi
i}\oint_{A_i}w\dd v&& \frac{\pa{\cal F}}{\pa S_i}=\oint_{B_i}w\dd v~.
\eea
$N_i$ is the classical $U(N)\rightarrow\prod_{i=1}^nU(N_i)$
symmetry breaking pattern, $\t$ is the $U(1)\subset U(N)$ coupling
constant \cite{civ} (which is the bare coupling constant of the YM
theory), the $\t_i$ are integers that describe generalized
theta-angles \cite{csw1}, $S_i$ are the corresponding gluino
condensates, and $\frac{\pa{\cal F}}{\pa S_i}$ are the dual
magnetic variables. $\hat B_i$ are the non-compact $B$-periods, i.e. $\hat 
B_i=B_i+B_n$ where $B_i$, the compact periods, and the non-compact period 
$B_n$, are defined in appendix \ref{perintegrals}.
These relations become obvious once we realize
that $\dd t/t$ and $w\dd v$ are related to the gauge theory and
matrix model resolvents. Namely
\be \label{rss1}
\frac{\dd t}{t} = {\rm Tr}_{\rm gauge\,\,theory} \left( \frac{\dd v}{v-\Phi} 
\right)
\ee
is the gauge theory resolvent \cite{cdsw,csw1}, whereas
\be \label{rss2}
w\dd v = 2{\rm Tr}_{\rm matrix\,\,theory} \left( \frac{\dd v}{v-M}
\right) = -\frac{1}{16\pi^2}\,{\rm Tr}_{\rm gauge\,\,theory} \left(
\frac{W_{\alpha} W^{\alpha} \dd v}{v-\Phi} \right)
\ee
\cite{dv1,cdsw}. Thus the M5-brane treats the gauge theory and
matrix model on a symmetric footing: the gauge theory curve is the
projection in the $t,v$-plane, whereas the matrix model curve is
the projection in the $t,w$ plane. The above relations are valid
on-shell. To go off-shell, we want to be able to vary the $S_i$'s
at will, keeping the periods of $\dd t/t$ fixed. From the M5-brane
point of view, this corresponds to deforming the way the Riemann
surface is embedded in ${\mathbb R}^5\times S^1$. Thus we look for
a continuous family of deformations of $\S$. However, we will find
that such deformations do not exist unless they are
non-holomorphic, which makes $t=t(v,\bar v)$ a function of both
$v$ and $\bar v$. This precisely parallels the discussion in the
gauge theory \cite{csw2}, where it was found that the one-form
$T$, which corresponds to our $\dd t/t$, had to be
non-holomorphic. We will analyze in detail this non-holomorphic
embedding, and we will find that it can be solved only if we
include also a log-normalizable anti-holomorphic deformation.

Notice that only $t$, and not $w$, will be non-holomorphic off-shell.
Indeed, the fact that $t$ parametrizes a cylinder, 
$t=\exp(-(x^6+ix^{10})/R)$ with
$x^{10}$ the compact M-theory direction with radius $R$, implies that the 
periods
of $\dd t/t$ have to be integers, and this gives an interesting new
purely geometrical
interpretation of the integrality of the periods of $T$ around the compact
cycles. Once we go off-shell we only change the periods of $w\dd v$ while 
keeping its
asymptotic behavior fixed. Under T-duality to the IIB description, the 
one-form $\dd t/t$ maps
to the flux three-form $H$, while the one-form $w\dd v$ maps to the complex 
structure
three-form $\Omega$. In this way the deformations of the Riemann-surface are 
in one-to-one
correspondence with complex structure deformations of the T-dual Calabi-Yau.

The main goal in this paper is to study the non-holomorphic deformations
of the M5-brane and to rederive the gauge theory superpotential from this 
perspective.
In particular we will encounter several different
definitions of the superpotential that all turn out to be equivalent.
The outline of this paper is as follows. In section~2 we will briefly review 
the
brane construction of the gauge theory and its lift to M-theory. We will 
also describe
the way in which the world-volume is twisted and see that this may provide a 
natural
framework to understand a peculiar auxiliary supermultiplet found in 
\cite{cdsw}.
In section~3 we discuss the non-holomorphic embedding of the M5-brane, and 
use this to
compute the expression for the superpotential proposed in \cite{wi2}. In 
section~4 we
consider the non-holomorphic geometries in some more detail, and in 
section~5 we discuss some
further aspects of the M5-brane world-volume theory, such as the 
``deconstruction''
interpretation of the superpotential. Several open problems are given at the 
end in
section~6.

%
%
%
%
%
%
%
\section{The on-shell M5-brane construction}\label{sec2}

It is well-known that an M5-brane with worldvolume ${\mathbb R}^4\times\S$ 
preserves
half the supersymmetry if $\S$ is holomorphically embedded in a 
2-dimensional
complex space, for example in ${\mathbb R}^3\times S^1$. The fluctuations of 
this Riemann surface 
in ${\mathbb R}^3\times S^1$ are then described by a single complex scalar 
superfield. In order to 
break
supersymmetry to ${\cal N}=1$ we need to consider a configuration where two
complex scalars are active, which happens for instance when $\S$ is a 
holomorphic two-cycle in
a Calabi-Yau $X$. In
fact, the requirements of meromorphy of the embedding and a choice of 
boundary
conditions at infinity are enough to determine the fivebrane geometry.
In this section we will review this on-shell
description.

\subsection{The type IIA picture}

It is useful to start from type IIA and then lift to M-theory. The
${\cal N}=2$ $SU(N)$ super-Yang Mills theory can be engineered with $N$
D4-branes suspended between two NS5-branes \cite{wi1}. Our notation will
be as follows. We denote the
coordinates along the worldvolume of the D4-branes by $x^0,x^1,x^2,x^3,x^6$,
and those of the NS 5-branes by $x^0,x^1,x^2,x^3,x^4,x^5$. The NS 5-branes 
are
at a finite distance from each other along the
$x^6$-direction. Classically, the positions of the D4-branes in
transverse space are
fixed, but quantum mechanically they are allowed to fluctuate, becoming
scalar fields on the brane
with fixed boundary conditions at the ends, where the NS 5-branes sit.
Together with $x^6$ and the
gauge field component along $x^6$, the scalar fields are parametrized by
$v=x^4+ix^5$, $w=x^7+ix^8$. The
classical rotation invariances along the $x^4,x^5$ and
$x^7,x^8,x^9$ directions,
$U(1)_{4,5}$ and $SU(2)_{7,8,9}$, respectively, correspond to the classical
$U(1)$ and $SU(2)$
R-symmetries of the 4-dimensional theory on the D4-branes.

To obtain a configuration with ${\cal N}=1$ supersymmetry with arbitrary 
superpotential for the
adjoint scalar field,
we consider the brane
construction in \cite{bo}, in which the first NS 5-brane extends
along $x^0,x^1,x^2,x^3,x^4,x^5$, whereas $N-1$
NS' 5-branes have worldvolume coordinates $x^0,x^1,x^2,x^3,x^7,x^8$.
The D4-branes between them
have worldvolume coordinates $x^0,x^1,x^2,x^3,x^6$.
(Constructions involving non-generic superpotentials were also given in
\cite{egk1,egk2,hoo,wi2}).
As shown in
\cite{bo}, adding a superpotential of the form
\be
W_{\sm{tree}}=\sum_{k=2}^N\m_k\Tr\F^k
\label{wtree}\ee
corresponds to bending the NS' 5-branes, allowing them to stretch in the
$(v,w)$ plane. Classically, the minima of \eq{wtree}
give the separation of the NS' 5-branes in the $v$-direction.
It will however be more convenient to view the NS' 5-branes
as a single NS 5-brane bent in the $w$-direction, rather than
$N-1$ NS' 5-branes bent in the $v$-direction, since this is
the appropriate configuration for finite values of the $\mu_k$ \cite{bo}.

\subsection{Lifting to M-theory}

The M-theory configuration \cite{wi1} consists of a single
M5-brane wrapped on a 2-cycle $\Sigma$ of a Calabi-Yau $X={\mathbb
R}^5\times S^1$. $\S$ will generically be a Riemann surface of
genus $N-1$. The world-volume of the M5-brane is thus
${\mathbb R}^4\times\Sigma$, and the total space is ${\mathbb R}^4\times 
X\times {\mathbb R}$;
as a separate problem one can also take the total space to be
${\mathbb R}^4\times X\times S^1$, so that one can T-dualize from IIA to the
IIB picture. This is however not quite the standard 4d gauge
theory, since it corresponds to having a compact $x^9$-direction.
The field theory lives on the ${\mathbb R}^4$ spanned by $x^0,\ldots,x^3$,
whereas $t,v,w$ are complex coordinates $v=x^4+ix^5$,
$w=x^7+ix^8$, $t=e^s=e^{-R^{-1}(x^6+ix^{10})}$ on $X$. They can be
used to describe the transverse fluctuations of $\Sigma$ embedded
in $X$.

As shown in \cite{wi1}, for ${\cal N}=2$ theories,
$\S$ is a Riemann surface given by an algebraic equation in the $(v,t)$
plane, the
${\cal N}=2$ Seiberg-Witten curve:
\be \label{pp1}
t^2-2P_N(v)\,t+\L^{2N}_{N=2}=0~,
\ee
where $P_N(v)$ is a polynomial of order $N$ in $v$. $t$ is related to the
more commonly used variable $y$, defined by
\be
y^2=P_N^2(v) -\L^{2N}_{N=2}~,
\ee
by $t=y+P_N(v)$. However, in the remainder of the paper, we will use $y(v)$ 
not to
denote the
variable defined by the above relation, but by the relation
\be
S_{N-n}^2(v)\, y^2=P_N^2(v) -\L^{2N}_{N=2}~,
\ee
where $S_{N-n}(v)$ contains all the double zeroes of $P_N^2(v) 
-\L^{2N}_{N=2}$.

The M-theory configuration corresponding to the IIA set-up with a
superpotential $W_{\sm{tree}}$ was
constructed in \cite{bo}. The left NS 5-brane corresponds to the asymptotic
region $v\rightarrow\infty$,
$t\sim v^N$, and the right NS 5-brane to $v\rightarrow\infty$,
$t\sim\L^{2N}_{N=2}v^{-N}$.
One imposes the following boundary conditions on $w$ at infinity:
\bea
w&\rightarrow& 2W_m'(v)\,\,\,\,\,\,{\mbox{as}}\,\,\,\,\,\,\,v
\rightarrow\infty,\,\,\,\,\,\,t\sim\L_{N=2}^{2N}v^{-N}\nn
w&\rightarrow&0\,\,\,\,\,\,\,\,\,\,\,\,\,\,\,\,\,\,\,\,\,\,{\mbox{as}}\,\,\,\,\,\,\,v
\rightarrow\infty,\,\,\,\,\,\,t\sim
v^N~,
\eea
where $W_m'(v)$ is a polynomial of order $m$ that corresponds to the field
theory tree level superpotential. Notice that here $m$ does not have to be
equal to $N-1$. Physically, the above means that we bend one of the two NS
5-branes in such
a way that the masses of open strings stretched between a straight D4 brane
and one of the two 5-branes is proportional to $W_m'(v)$, in agreement
with the classical properties of the field theory deformed by the
superpotential (\ref{wtree}).
As we will see, these two regions correspond to the two ``branches",
``quantum"
and ``classical", respectively, in \cite{csw2}.

There is an alternative choice of boundary conditions, but this only
corresponds to interchanging the two NS 5-branes. More generally, we could
impose $w\rightarrow aW'_m(v)$ on one branch, and $w\rightarrow bW'_m(v)$ on
another, with $a-b=2$, but that will not affect any of the F-term physics.
It might have some effect on non-holomorphic terms in the 4d effective 
action
but we will not explore that here.

In the analysis of \cite{bo}, the fact that the form of the ${\cal
N}=2$ curve remains unchanged under the breaking to ${\cal N}=1$
means that the embedding coordinate $t$ still satisfies equation
\eq{pp1}. Breaking to ${\cal N}=1$, however, does require this
curve to factorize, as we will now see. As for $w$, which was zero
in the ${\cal N}=2$ case -- its form can be fixed from the
requirement that it is a rational function of $t$ and $v$
satisfying the above boundary conditions. The analysis then leads
to the following conclusions (for details we refer to \cite{bo}).
The requirement that $w$ be a rational function of $t,v$ with no
other poles than at infinity, implies that it can be written in
the following form:
\be
w(t,v)=N_m(v) +\frac{H_{m-n}(v)}{ S_{N-n}(v)}(t-P_N(v)),
\ee
where $N_m(v)$ and $H_{m-n}(v)$ are as yet arbitrary polynomials of the 
order
indicated, $P_N(v)$ is the polynomial appearing in the ${\cal N}=2$ curve, 
and
$S_{N-n}(v)$ is the set of double roots in the factorization formula of the
Seiberg-Witten curve:
\be
P_N^2(v)-\L^{2N}_{N=2}=S_{N-n}^2(v)Q_{2n}(v)~,
\ee
so that $S_{N-n}$ and $Q_{2n}$ are given by
\bea
S_{N-n}(v)&=&\prod_{i=1}^{N-n}(v-s_i)\nn
Q_{2n}(v)&=&\prod_{i=1}^{2n}(v-q_i)
\eea
where all $q_i$'s are different. Using the expression for $t$ in 
(\ref{pp1}), $w$ can also
be rewritten as
\be
w(t_\pm(v),v)=N_m(v)\pm H_{m-n}(v)\sqrt{Q_{2n}(v)}~,
\ee
depending on whether we choose the plus or the minus solution. These two
different choices correspond to the two different asymptotic regions.

One now has to impose the boundary conditions on $w$. The first boundary
condition, $t=t_-(v)\sim v^N$ as $v\rightarrow\infty$, completely fixes
$N_m(v)$:
\be \label{au1}
N_m(v)=[H_{m-n}(v)\sqrt{Q_{2n}(v)}]_+~,
\ee
where $[Q(v)]_+$ is the part of $Q$ with non-negative powers of $v$, in a
power series expansion around $v=\infty$. Substituting this in $w$, we get 
in the
second asymptotic region
\be
w(t_+(v),v)=2[H_{m-n}(v)\sqrt{Q_{2n}(v)}]_+ +{\cal
O}\left(\frac{1}{v}\right)
\ee
as $v\rightarrow\infty$, $t=t_+(v)\sim v^{-N}$. The second boundary
condition thus amounts to
\be \label{au2}
N_m(v) = [H_{m-n}(v)\sqrt{Q_{2n}(v)}]_+=W_m'(v)~.
\ee
The two equations (\ref{au1}) and (\ref{au2}) can be rephrased as
$N_m=W_m'$ and $\tilde{f}_{m-1}+N_m^2-Q_{2n}H_{m-n}^2=0$ for some polynomial
$\tilde{f}_{m-1}$
of order $m-1$, and $m$ can in general be different from $N-1$.

Summarizing, the world-volume of the M5-brane is ${\mathbb R}^4\times\S$
where $\S$ is a Riemann surface embedded in ${\mathbb R}^5\times
S^1$ with complex coordinates $t,w,v$. The Riemann surface is
defined by the following embedding:
\bea\label{tw}
t^2-2P_N(v)t+\L^{2N}_{N=2}&=&0\nn
w^2-2W'_m(v)w-\tilde{f}_{m-1}(v)&=&0~,
\eea
which indeed gives a one-dimensional complex surface. Notice that $\S$ can
be viewed as the intersection of two different Riemann surfaces, one in the
$t-v$ plane defined by the first equation, and the other in the $w-v$ plane,
defined by the second equation. This ``doubling'' precisely parallels the 
field
theory behavior \cite{cdsw}.

Furthermore, ${\cal N}=1$ supersymmetry imposes factorization formulas on 
this
Riemann surface, so that its genus is reduced from $N-1$ to $n-1$:
\bea \label{jj1}
P_N^2(v)-\L^{2N}_{N=2}&=&S_{N-n}^2(v)Q_{2n}(v)\nn
W_m'(v)^2 + \ti f_{m-1} &=& H_{m-n}^2(v) Q_{2n}(v)~.
\eea
These are the requirements that there exist meromorphic functions on the
Riemann surface with the prescribed boundary conditions.

So we see that, except at the points where $v$ blows up, one can
think of $v$ and
\be
y=\sqrt{Q(v)}
\ee
as coordinates on the Riemann surface, which has genus $n-1$, with branch 
cuts
between the $2n$ zeroes $q_i$ of $Q(v)$. We will follow the notation in
\cite{csw2}, where a point is denoted by $p$ if $(v,y)=(p,\sqrt{Q(p)})$ and 
by
$\ti p$ if $(v,y)=(p,-\sqrt{Q(p)})$. We will also decompose $Q(v)$ as
\be \label{jjj1}
Q(v) = G_n^2(v) + f_{n-1}(v).
\ee

At this point it is clear that $t$ and $w$ have precisely the same
properties as the gauge theory and matrix model resolvents \cite{dv1,cdsw},
where
\be \label{resoo}
T(v) = \frac{\dd t}{t} = {\rm Tr}_{\rm gauge\,\,theory} \left( \frac{\dd 
v}{v-\Phi} \right)
=\dd\log(P_N+y)
\ee
is the gauge theory resolvent, whereas
\bea \label{reso}
R(v)&  =&  w\dd v = 2\,{\rm Tr}_{\rm matrix\,\,theory} \left(
\frac{\dd v}{v-M} \right) = -\frac{1}{16\pi^2}\,{\rm Tr}_{\rm
gauge\,\,theory} \left( \frac{W_{\alpha} W^{\alpha} \dd v}{v-\Phi}
\right) \nn & = & W_m'(v) - \sqrt{W_m'(v)^2+\tilde{f}_{m-1}(v)}
\eea
is the matrix model resolvent.
Notice that the two branches, classical and quantum, of the Riemann
surface, arise naturally as the two asymptotic regions of the 5-brane.

So far the discussion was on-shell. Indeed, fixing the boundary conditions
$W'_m(v)$ at infinity completely fixes all the remaining coefficients, and 
so
there is no room for holomorphic deformations of the brane configuration.
In the next section we will
see how to define deformations of $y$ so our discussion can parallel the
deformations of the gauge theory where one turns on expectation values for 
the
gluino condensates. We first discuss a few aspects of the twisted 
world-volume
theory.

\subsection{The twisted $(2,0)$ theory}

The world-volume theory of M5-branes is quite complicated due to
the presence of a self-dual tensor field. For general curved
world-volumes embedded in eleven-dimensional supergravity, the
world-volume theory for a single M5-brane was constructed in
\cite{m5a,m5b}. A simple action for a flat M5-brane in flat eleven
dimensions is given in \cite{m5c}, although this action also
depends on the anti-self dual part of the tensor field. The latter
paper also contains a useful summary of the various symmetries of
the theory.

The standard $(2,0)$ theory has sixteen supercharges that
transform as $({\bf 4},{\bf 4})$ under $SO(5,1)\times SO(5)_R$,
the product of the Lorentz group and the $SO(5)$ R-symmetry group.
However, once we wrap the M5-brane on a non-trivial Riemann
surface, the world-volume theory is automatically twisted
\cite{bsv}. The twist is such that the transverse scalars, which
for a flat brane live in a trivial bundle over the world-volume,
become sections of the normal bundle over the Riemann surface. To
describe the twist, we notice that the Lorentz group is reduced
from $SO(5,1)$ to $SO(3,1)\times SO(2)$ once we wrap the fivebrane
on a non-trivial surface. Twisting the theory means that we embed
the $SO(2)$ part of the broken Lorentz group in the $SO(5)$
R-symmetry group. Since one of the five scalars is always trivial
($x^9$ in our notation), we need to embed $SO(2)$ in a
$SO(4)=SU(2)\times SU(2)$ subgroup of $SO(5)_R$.

To determine the quantum numbers after the twisting, we first write the 
quantum
numbers of the scalars and supercharges in terms of $SU(2)\times SU(2)\times 
U(1)\times
SU(2)_L\times SU(2)_R$, where the first two $SU(2)$'s represent the Lorentz 
group
$SO(3,1)$\footnote{Rather, $SU(2)\times SU(2)$ represents the Euclidean 
Lorentz group,
but we trust that this will not cause any confusion.}, the $U(1)\sim SO(2)$ 
represents
the internal part of the Lorentz group, and the last two $SU(2)$'s represent
the $SO(4)\subset SO(5)$ subgroup of the R-symmetry group that plays a role 
in the
twisting. Before the twisting, the scalars and supercharges transform as
\be
\begin{array}{|c|c|} \hline
\Phi &
[( {\bf 1},{\bf 1})_0, ({\bf 2},{\bf 2})]\oplus
[( {\bf 1},{\bf 1})_0, ({\bf 1},{\bf 1})] \\ \hline
Q & [( {\bf 2},{\bf 1})_1, ({\bf 2},{\bf 1})]\oplus
[( {\bf 2},{\bf 1})_1, ({\bf 1},{\bf 2})] \\ \hline
\bar{Q} & [( {\bf 1},{\bf 2})_{-1}, ({\bf 2},{\bf 1})]\oplus
[( {\bf 1},{\bf 2})_{-1}, ({\bf 1},{\bf 2})] \\ \hline
\end{array}
\ee
where $[({\bf j_1},{\bf j_2})_m, ({\bf j_3},{\bf j_4})]$ contains
the quantum numbers $(j_1,j_2,m,j_3,j_4)$ under $SU(2)\times
SU(2)\times U(1) \times SU(2)_L \times SU(2)_R$, the first two
$SU(2)$'s being the Lorentz group and the last two the R-symmetry.

The $U(1)$ that rotates $\log t$ but not $w$ is the $U(1)$ that is 
diagonally embedded in
$SU(2)_L \times SU(2)_R$, whereas the $U(1)$ that rotates $w$ but not $\log 
t$ is
the off-diagonal one. Therefore, the ${\cal N}=2$ twist is the one where the
$U(1)$ part of the Lorentz group is diagonally embedded in $SU(2)_L\times 
SU(2)_R$,
whereas the ${\cal N}=1$ twist is the one where the $U(1)$ is embedded in 
$SU(2)_L$
only. After the ${\cal N}=2$ twist the fields transform as
\be
\begin{array}{|c|c|} \hline
\Phi &
( {\bf 1},{\bf 1})_0\oplus
( {\bf 1},{\bf 1})_0\oplus
( {\bf 1},{\bf 1})_0\oplus
( {\bf 1},{\bf 1})_2\oplus
( {\bf 1},{\bf 1})_{-2}
\\ \hline
Q &
( {\bf 2},{\bf 1})_2 \oplus
( {\bf 2},{\bf 1})_2 \oplus
( {\bf 2},{\bf 1})_0 \oplus
( {\bf 2},{\bf 1})_0 \\ \hline
\bar{Q} &
( {\bf 1},{\bf 2})_0 \oplus
( {\bf 1},{\bf 2})_0 \oplus
( {\bf 1},{\bf 2})_{-2} \oplus
( {\bf 1},{\bf 2})_{-2} \\ \hline
\end{array} ~.
\ee
From this we see that half of the supersymmetries transform as a scalar
on the Riemann surface, and these supersymmetries give rise to the unbroken
${\cal N}=2$ in four dimensions. In addition, $\log t$ transforms as a
section of ${\cal O}(-2)$ and $w$ as a section of ${\cal O}(0)$. This is
consistent with the fact that the normal bundle of the Riemann surface
in $\mathbb R^5 \times S^1$ should have degree $-2$.

If we embed $U(1)$ in $SU(2)_L$, the scalar fields and supercharges 
transform
as
\be
\begin{array}{|c|c|} \hline
\Phi &
( {\bf 1},{\bf 1})_0\oplus
( {\bf 1},{\bf 1})_1\oplus
( {\bf 1},{\bf 1})_1\oplus
( {\bf 1},{\bf 1})_{-1}\oplus
( {\bf 1},{\bf 1})_{-1}
\\ \hline
Q &
( {\bf 2},{\bf 1})_2 \oplus
( {\bf 2},{\bf 1})_1 \oplus
( {\bf 2},{\bf 1})_1 \oplus
( {\bf 2},{\bf 1})_0 \\ \hline
\bar{Q} &
( {\bf 1},{\bf 2})_0 \oplus
( {\bf 1},{\bf 2})_{-1} \oplus
( {\bf 1},{\bf 2})_{-1} \oplus
( {\bf 1},{\bf 2})_{-2} \\ \hline
\end{array} ~.
\ee
Now both $\log t$ and $w$ transform as sections of ${\cal O}(-1)$. Only 
$1/4$
of the supersymmetries transform as a scalar, and these yield the surviving
${\cal N}=1$ generators in four dimensions.
The R-symmetry of the resulting
${\cal N}=1$ field theory acts as (minus) the internal $U(1)$ piece of the 
Lorentz
group before twisting. In particular, $w$ and $\log t$ have no $R$-charge,
whereas the one-form $\dd v$ has R-charge 2, and $Q$ has R-charge $-1$.
In terms of the $SU(2)\times SU(2) \times U(1) \times U(1)_L \times U(1)_R$ 
subgroup of
$SU(2)\times SU(2) \times U(1) \times SU(2)_L \times SU(2)_R$, the charges 
of
$\log t$, $w$ and the unbroken ${\cal N}=1$ generator $Q_{\alpha}$ are
\be
\begin{array}{|c|c|} \hline
\log t &
[( {\bf 1},{\bf 1})_0, (-1,-1)]
\\ \hline
w &
[( {\bf 1},{\bf 1})_0, (-1,+1)]
\\ \hline
Q_{\alpha} &
[( {\bf 2},{\bf 1})_1 , (-1,0)]
\\ \hline
\end{array} ~.
\ee
We normalized the eigenvalues of $U(1)$ in such a way that they are always 
integer, i.e.
they are twice the spin.
We also pick one of the other three supersymmetry generators (which is 
broken once
we wrap the five brane), denoted by $\hat{Q}_{\alpha}$,
with quantum numbers $[( {\bf 2},{\bf 1})_1 , (0,-1)]$. Though the 
corresponding
symmetry is broken, we can still use it to build a supermultiplet, which we 
write
in terms of an auxiliary superspace with anti-commuting coordinates 
$\eta^{\alpha}$.
Using the quantum numbers of $\hat{Q}_{\alpha}$ we find that an example of 
such a multiplet is
\be
F=w + \chi_{\alpha} \eta^{\alpha} + \frac{\partial \log t}{\partial v} 
\eta^{\alpha}
\eta_{\alpha}
\ee
with some anti-commuting degree of freedom $\chi_{\alpha}$. if we compute 
the
A-periods (\ref{per1}) of the differential $F\dd v$, we obtain a 
four-dimensional supermultiplet
of the form
\be
\frac{1}{2\pi i}\oint_{A_i} F\dd v  = S_i + w_{\alpha} \eta^{\alpha}
+ N_i \eta^{\alpha} \eta_{\alpha}
\ee
which has precisely the form of the auxiliary supermultiplet introduced in 
\cite{cdsw}.
The full effective action can be written as an integral over auxiliary 
superspace
of a holomorphic function of this auxiliary supermultiplet. Thus, this 
auxiliary structure
seems to have a direct origin in the supersymmetry of the M5-brane. It would 
be interesting
to work out the twist and the corresponding KK reduction in more detail, as 
this would no
doubt shed further light on the structure of the four-dimensional effective 
action.

\section{The off-shell superpotential}\label{sec3}

In \cite{wi1}, Witten defined a superpotential in terms of the M5-brane
coordinates that reproduces the on-shell value of the effective 
superpotential in the
gauge theory. Holomorphic configurations are minima of this superpotential.

The aim of this section is to show that this superpotential not only gives 
the
correct on-shell result, but can also be evaluated off-shell, and its 
off-shell
value reproduces the Dijkgraaf-Vafa result in terms of the gluino
condensates.

The precise definition of the superpotential is actually quite subtle, 
because the
Riemann surface we are considering is non-compact. However, before we turn 
to a discussion
of the superpotential we first need to describe the off-shell geometry of 
the M5-brane.

\subsection{Off-shell deformations of the M5-brane}
\label{defo}

The on-shell configuration for the five-brane, as given in equations 
(\ref{tw}), (\ref{jj1}) and
(\ref{jjj1}), reads
\bea\label{twa}
t^2-2P_N(v)t+\L^{2N}_{N=2}&=&0\nn
w^2-2W'_m(v)w-\tilde{f}_{m-1}(v)&=&0~,
\eea
where
\bea
P_N^2 - \Lambda^{2N} & = & S^2_{N-n} (G_n^2 + f_{n-1})
\label{eqq1}
\\
(W'_m)^2 + \tilde{f}_{m-1} & = & H^2_{m-n} (G_n^2 + f_{n-1}) .
\label{eqq2}
\eea
Both $t$ and $w$ can be viewed as functions on the underlying
Riemann surface $y^2=G_n^2 + f_{n-1}$. We are interested in
varying this underlying Riemann surface, so that the glueball
degrees of freedom $S_i\sim {\rm Tr}_{U(N_i)} ({\cal W}^2)$
acquire a non-trivial expectation value. According to
(\ref{per1}), this is the same as changing the $A$-periods of the
differential $w\dd v$, which can be accomplished by varying the
function $f_{n-1}$. However, we should also specify the embedding
of this new Riemann surface in $\mathbb R^5 \times S^1$, i.e. give
new functions $t$ and $w$ on this Riemann surface. Since the
glueball fields $S_i$ are chiral superfields, one might be
inclined to believe that the relevant deformations of $t$ and $w$
should be holomorphic. As we will now argue, this turns out to be
incorrect.

If $t$ and $w$ were still holomorphic after the deformation, one
might believe that they would still obey equations of the form
(\ref{eqq1}) and (\ref{eqq2}). A simple count of the number of
free parameters shows that this cannot work. Once we specify
$W'_m$ and the $n$ $A$-periods of $w\dd v$, which we trade off for
the coefficients in $f_{n-1}$, the number of free parameters
remaining in equations (\ref{eqq1}) and (\ref{eqq2}) is
$N+(N-n)+n+m+m-n=2N+2m-n$. The number of equations is $2N+2m$,
which is larger than the number of free parameters, and the system
will generically have no solution.

Of course, there could exist holomorphic deformations that don't
preserve the factorized form (\ref{eqq1}) and (\ref{eqq2}). A
naive counting argument shows that this is unlikely as well. If we
consider a compact Riemann surface $\Sigma$ of genus $g$ embedded
in a Calabi-Yau of complex dimension $d$, then a straightforward
application of the index theorem yields that ${\rm dim}\,H^0(\Sigma,
{\cal N}) -{\rm dim}\,H^1(\Sigma, {\cal N}) = (d-3)(1-g)$, with
${\cal N}$ the normal bundle, so as long as ${\rm
dim}\,H^1(\Sigma,{\cal N})=0$ the number of holomorphic sections of
the normal bundle is simply $(d-3)(1-g)$. This is quite naive for
many reasons, first of all in our case the Riemann surface is
noncompact, there may in general be obstructions to lift a
holomorphic section of the normal bundle to a finite holomorphic
deformation, and $H^1$ may be nonzero. Nevertheless, if we apply
this result to the present case, we see that there are no
holomorphic deformations, as $d=3$.

If we ignore the coordinate $t$ for the moment, then a count of
parameters in equation (\ref{eqq2}) shows that this equation does
generically have a solution for arbitrary $f_{n-1}$. Thus, there
is no problem defining a holomorphic function $w$ on the Riemann
surface, all problems are caused by the coordinate $t$.

To resolve this issue, it is helpful to slightly change
perspective and to think of the one-forms $w\dd v$ and $\dd t/t$
instead of the functions $w$ and $t$. These have an interpretation
as matrix model and gauge theory resolvents as in equations
(\ref{rss1}) and (\ref{rss2}), and it should be possible to
determine them off-shell. This was indeed done in \cite{csw2}.
Off-shell, $w\dd v$ is some differential with periods around the
cycles of the Riemann surface that are directly related to the
glueball vevs. Therefore, we expect that once we go off-shell, the
relation (\ref{eqq2}) is still satisfied, with ${f}_{n-1}$
parametrizing the deformations. In other words, in the $v,w$ plane
going off-shell simply corresponds to making a holomorphic change
of the Riemann surface.

The situation regarding $t$ is more complicated. In a vacuum of
the theory, the holomorphic one-form $\dd t/t$ has integer periods
around all compact cycles of the Riemann surface. The fact that
the periods are integer means that $\log t$ is a well-defined
function with values in ${\mathbb R}\times S^1$ on the Riemann
surface (or put differently, this follows from the fact that
$H^0(\Sigma,{\mathbb C}^{\ast}) \equiv H^1(\Sigma,{\bf Z})$). Once we go
off-shell by a geometric deformation in the $w,v$ plane, there no
longer is a holomorphic one-form with integer periods around all
cycles. However, there is a non-holomorphic closed one-form, which
is defined in such a way that it has the same integer periods
around both the A and B cycles as $\dd t/t$ had in the vacuum of
the theory, and the same pole structure at infinity. We will
continue to call this one-form $\dd t/t$. It is indeed appropriate
to call the form $\dd t/t$, because there still will be a global
${\mathbb R}\times S^1$ valued function $\log t$ associated to
$\dd t/t$, and we will take this function to define the embedding of
the Riemann surface in the $t$-direction. However, this embedding
will no longer be holomorphic.

To summarize: to go off-shell we need to perform a holomorphic deformation
in
the $w,v$-plane, and a non-holomorphic deformation in the $t$-direction,
defined
by a non-holomorphic one-form with integer periods.

The non-holomorphic one-form $\dd t/t$ is denoted by $T_0$ in
\cite{csw2}.
It is off-shell no longer equal to the resolvent of the gauge theory.
The latter is given by a holomorphic one-form $T$, which has the same
periods around the A-cycles as $T_0$, but non-integer B-periods. On-shell,
$T$ and $T_0$ coincide. We will discuss the fivebrane interpretation of
$T$ in section~\ref{reskin}.

Another perspective on the five-brane is that it represents a pure
geometrical dual of the setup of \cite{civ}. In this setup one
considers a noncompact Calabi-Yau with either wrapped branes or
with a non-trivial three-form flux with integer periods.
Quantities computed using this non-compact Calabi-Yau can be
computed in terms of a Riemann surface. The three-form flux
reduces to a one-form on this Riemann surface with integral
periods. This is exactly the same one-form as $\dd t/t$. Therefore
we see that after dualizing to M-theory, the flux becomes purely
geometrical and encodes the embedding of the fivebrane in the
ambient space in the $t$-direction. The periods of the flux become
the winding numbers of the ${\mathbb C}^{\ast}$ variable $t$ around the
corresponding cycles.

\subsection{Computing the superpotential}

Now that we have learned how to deform the Riemann surface, we
would like to consider the off-shell computation of the
superpotential for the fivebrane configuration, leaving a detailed
analysis of the non-holomorphic embedding $t$ for the next
section. Obviously, we expect to recover the standard expression
for the superpotential, but we would like to see how this comes
about.

We will use the notation in \cite{wi1}, where one thinks of
$\Sigma$ as a surface with a map $\F:\Sigma\rightarrow X$. We will
take this to be an embedding, although our considerations should
generalize as in \cite{wi1}. This map is described by functions
$\f^i(\l)$, where $\l$ are coordinates on $\Sigma$.


The expression that Witten postulated in \cite{wi1} for the
superpotential for M5 branes wrapping a two-cycle is
\be
W(\Sigma)-W(\Sigma_0) =\frac{1}{2\pi i}
\int_B\Omega \label{superpotential}\ee where $B$ is a 3-chain
interpolating between $\Sigma$ and $\Sigma_0$, and $\Sigma_0$ is
an arbitrary reference surface in the same homology class as
$\Sigma$. In our coordinates, the holomorphic 3-form $\Omega$ is
$\Omega=\O_{ijk}\dd\f^i\wedge\dd\f^j\wedge\dd\f^k=R\dd v\wedge\dd
w\wedge\dd t/t$. As a basis for one-forms $\dd\phi^i$ we will use
$\dd v,\dd w,\dd\log t$.


We will choose coordinates such that $v$ is a good function on either sheet
of the Riemann surface, except at infinity where it has a pole. Minimizing 
the
superpotential then amounts to finding suitable meromorphic functions 
$(t,w)$
on this surface. Notice that from the 5-brane perspective there is nothing
particular about this choice, and one could just as well interchange the 
roles
of $v$ and $w$.

Before proceeding, we point out that the definition of the superpotential
in (\ref{superpotential}) requires $\Sigma_0$ and $\Sigma$ to have the
same asymptotics, otherwise a certain regularization at infinity is
required. In particular, this implies that (\ref{superpotential}) can
only be used to study the variation under normalizable deformations of
the curve (though we will also allow log-normalizable deformations in the
remainder, but the final expression for the superpotential will be finite 
when
we send the cutoff to infinity even off-shell). Of course, we are also
interested in the variation of the
superpotential under non-normalizable deformations of the curve, such
as those associated to variations of the coefficients in the superpotential.
To determine this variation, we need to supplement (\ref{superpotential})
with some additional information, otherwise we could choose arbitrarily
different $\Sigma_0$ for each non-normalizable deformation and the
result for the superpotential would be arbitrary.

Note also that if $\Sigma$ were compact, the superpotential would
be zero for any holomorphic $\Sigma$. The fact that we get a non-trivial
superpotential as a function of holomorphic variables is due to
the non-compactness of $\Sigma$. (Another way in which this can
happen is if there is a global obstruction to certain holomorphic
deformations, see e.g. \cite{kach}).

We now first consider the extrema of the superpotential
(\ref{superpotential}). If we perform a generic variation of the
superpotential we obtain
\be
2\pi i\,\d W=3
\int_\Sigma\Omega_{ijk}\d\f^i\dd\f^j\wedge\dd\f^k=0
\ee
which we rewrite as
\be
\int_\Sigma\Omega_{ij}\d\f^i\wedge\dd\f^j=0
\label{minimum}\ee
where $\Omega_{ij}$ is a reduced 1-form on the Riemann surface,
$\O_{ij}=\O_{ijv}\dd v$, and $i$ runs
over $(\log t,w)$.

On $\S-\{P,Q\}$, this implies that $w$ and $t$ should holomorphic
functions of $v$. This is already enough to find the form of $w$ and
$t$, once we impose suitable boundary conditions on $w$ and $t$,
and assume the underlying Riemann surface is hyperelliptic.

To see this, we notice that any holomorphic
function $u$ on the hyperelliptic Riemann surface
$y^2 = G_n^2(v) + f_{n-1}(v)$ can be written as $u=A(v) + y B(v)$, with
$A$ and $B$ meromorphic functions of $v$. Clearly, $u$ satisfies
\be
(u-A(v))^2 = B^2(v) (G_n^2(v) + f_{n-1}(v))
\ee
and after multiplication with a suitable polynomial we see that
$u$ must satisfy an equation of the form
\be \label{ta1}
C(v) u^2 + D(v) u + E(v) =0
\ee
with polynomials $C,D,E$ with no common factors.
In addition, there must exist a polynomial
$F$ so that
\be \label{ta2}
D^2(v) - 4 C(v) E(v) = F^2(v) ( G_n^2(v) + f_{n-1}(v) ).
\ee
If $C(v)$ has a zero for some finite value of $v$, then at that value
of $v$ at least one of the two solutions $u$ of (\ref{ta1}) would go
to infinity, which contradicts the boundary condition that $u$ should
be holomorphic in $\Sigma-\{P,Q\}$. Therefore, $C(v)$ has to be a constant
and can be taken equal to one. The boundary conditions on $t$ and
$w$ together with the constraint (\ref{ta2})
then completely fix the form of the remaining polynomials $D(v)$ and
$E(v)$ for both $t$ and $w$.
Notice that $t$ should not be zero anywhere on $\S-\{P,Q\}$ and
therefore one immediately sees that for $t$ the polynomial $E$ should
be a constant. A zero of $t$ is not acceptable because $t$ was related
to $x^6$ and $x^{10}$ by $t=\exp(-R^{-1}(x^6+i x^{10}))$.

The final result for $t$ and $w$ is summarized in \eq{tw},
(\ref{eqq1}) and (\ref{eqq2}).

To obtain the superpotential as a function of glueball superfield
vacuum expectation values, we should give an expression for it
which is valid off-shell, i.e. once we turn on the
deformations discussed in section~\ref{defo}.

After a partial integration, and after dropping the contribution from
$\Sigma_0$ which is just a constant, we obtain
\be \label{pint}
W_{\sm{eff}}=\frac{R}{2\pi i}\int_\S\frac{\dd t}{ t}\wedge
w\dd v= \frac{R}{2\pi i}\int_\S \eta^1 \wedge \eta^2~,
\ee
where we defined the two one-forms
\be
\eta^1 = w \dd v, \qquad \eta^2 = \frac{\dd t}{t} .
\ee

It is important to pause for a moment here to discuss the partial
integration.
In general, on a non-compact Riemann surface, one has to determine
what one keeps fixed at infinity (see e.g. the discussion in \cite{av}
where the variable describing the size of disc instantons was kept
fixed). Here, what we would like to keep fixed is the coefficients
in the tree-level superpotential, which determine the asymptotics of
$w$. Thus, we keep $w$ fixed at infinity, and that is why we extract
$w$ in (\ref{pint}), rather then e.g. $\log t$.

The deformations can be parametrized by the coefficients in $f_{n-1}(v)$,
that appears in the defining equation
\be y^2 = G_n^2(v) + f_{n-1}(v)
\ee
for the Riemann surface. The polynomial $f_{n-1}(v)$ can be expanded as
\be\label{fexpansion}
f_{n-1}(v)=\sum_{i=0}^{n-1}f_i\,v^i~,
\ee
where the coefficients $f_i$ with $i=0,\ldots,n-2$ correspond to
normalizable deformations, and the coefficient $f_{n-1}$ to a
log-normalizable deformation\footnote{We use the same symbol for the 
polynomial
$f_{n-1}(v)$ and its coefficients, $f_i$, but hopefully this will not cause
any confusion.}. Equivalently, we can use the
A-periods of $\eta^1=w\dd v$ to parametrize the deformation,
\be
S_i = \frac{1}{2\pi i}\oint_{A_i}w\dd v ~,
\ee
where as usual the A-cycles are compact and run around the branch cuts of
$y$
in the $v$-plane as determined by the hyperelliptic form $y^2=G_n^2 +
f_{n-1}$ of $\Sigma$, see appendix \ref{perintegrals}. The A-periods of
$\eta^2=\dd t/t$ are also integers,
\be\label{periodsT}
2\pi iN_i = \oint_{A_i}\frac{\dd t}{t}~,
\ee
which are the ranks of the gauge groups that remain classically
unbroken in a particular classical minimum of the superpotential.
This can easily be seen by taking the classical limit
$\Lambda_{N=2}\rightarrow 0$ of the M5 brane configuration. In
that case,
\be
P_N(v)\rightarrow\prod_{i=1}^n(v-c_i)^{N_i}
\ee
and so the periods are equal to $2\pi iN_i$.

The A-cycles are dual to noncompact B-cycles. We know from the special
geometry relations of Calabi-Yau manifolds that
\be
\frac{\partial {\cal F}_0}{\partial S_i} =\frac{1}{2\pi i}
\oint_{B_i}w\dd v ~,
\ee
where ${\cal F}_0$ is the prepotential. If we in addition for the
moment assume that the integer periods of $\dd t/t$ around the
compact B-cycles all vanish, then the integrals of $\dd t/t$ along
all non-compact B-cycles are all identical,
\bea
2\pi i\tau &=&\int_{B_n}\frac{\dd t}{t}\nn
0&=&\oint_{B_i}\frac{\dd t}{ t}~,
\eea
$i=1,\ldots,n-1$, where $\tau$ will have the interpretation as the gauge
coupling.

At this point we can invoke the Riemann bilinear identities, and rewrite
(\ref{pint}) as
\be
W_{\sm{eff}}=R\sum_{i=1}^n\left(N_i\frac{\pa{\cal F}_0}{\pa
S_i}-2\pi i\t S_i\right)= R\sum_{i=1}^nN_i\frac{\pa{\cal F}_0}{\pa S_i}
-RNS\,\log\frac{\L_{N=2}}{\L_0}~, \label{superpot}\ee which is
precisely the Dijkgraaf-Vafa answer. The scale $\L_0$ appears
because the integral of $\dd t/t$ around non-compact cycles needs
to be regularized, see appendix~A.

Actually, once we accept (\ref{pint}) and once we identify $\dd
t/t$ with the one-form that one obtains by reducing the three-form
flux on the dual type IIB side, we see that (\ref{pint}) is
nothing but the reduction of the type IIB superpotential $\int
H\wedge \Omega$ to the Riemann surface. The minimalization of
(\ref{superpot}) from this point of view has already been
discussed in \cite{cv}. They showed that at the minimum of
(\ref{superpot}) there exists a meromorphic function with an
$N$-th order pole at $\ti P$ and an $N$-th order zero at $P$. Of
course, this meromorphic function is nothing but $t$, and at the
minimum we recover the holomorphic fivebrane configuration of
M-theory.

This discussion is somewhat imprecise, because (\ref{pint}) is still
not quite well-defined as it stands, and because there is in principle
no reason why the periods of $\dd t/t$ around the compact B-cycles should
vanish. A more precise discussion and the interpretation of the periods
of $\dd t/t$ around the B-cycles is given in \cite{csw1,csw2}, see in
particular section~7.1 of \cite{csw2}. There, a regulated version of
(\ref{pint}) is defined; as it stands, (\ref{pint}) is problematic
because both $\eta^1$ and $\eta^2$ have poles at $P,\ti P$. To make it
well-defined, one moves the poles of $\eta^1=\dd t/t$ to nearby
points $P',\tilde{P}'$. Next one applies a suitable version
of the Riemann bilinear identities for closed differentials with poles,
of the type discussed in section III.3 of \cite{farkaskra}. This then
gives a precise regulated definition of (\ref{pint}), and one can again
show quite generally that at the extremum $\dd t/t$ becomes a meromorphic
differential. The only effect of the periods, $\t_i$, of $\dd t/t$ around 
the
compact B-cycles is an extra term
\be
W'_{\sm{eff}}=2\pi iR \sum_{i=1}^{n-1} \t_i S_i
\ee
in (\ref{superpot}).

To see more directly that on-shell $\dd t/t$ is the gauge theory
resolvent we manipulate the expression in (\ref{pint}) as follows.
We first write it as
\be\label{wlogt}
\frac{R}{2\pi i} \int_\S w\,\bar{\pa} \log t \, \dd v \dd
\bar{v}.
\ee

Naively, this expression vanishes, as on-shell $\log t$ is a holomorphic
function of $v$. However, one has to be careful, as $t$ has poles at 
infinity. We can rewrite this expression as follows:
\bea
W_{\sm{eff}}&=&-\frac{R}{2\pi i}\int_\S\dd(w\log t\dd v)=-\frac{R}{2\pi i} 
\oint_Pw\log t\dd v\nn
&=&-2R\,{\mbox{Res}}_{v=\infty}\left(\log t(W_m'(v)+{\rm 
subleading})\right)\nn
&=&2R\,{\rm Res}_{v=\infty}\left( \frac{\pa \log t}{\pa v} W_m(v) \right)
\eea
where we assume that $P$ is the point where $t\sim v^N$, $w\sim 2W'_m(v)$.
In order for this to be the right answer, we need that $\dd \log
t$ is proportional to the gauge theory resolvent, see also
(\ref{resoo}), so that this equation reduces to the classical
superpotential evaluated at the right quantum vacuum.

Thus we see that these two ways to interpret (\ref{pint}) give two
equivalent answers for the superpotential, one off-shell as in
Dijkgraaf-Vafa, the other directly on-shell in terms of the gauge
theory resolvent. Interestingly, the computations appearing here
bear a close resemblance to those that appear in the study
\cite{is1,is2} of the integrable models that are needed to write
down the superpotentials of these theories once they are
compactified on $S^1$. At this point this seems somewhat
coincidental, as the degrees of freedom describe off-shell
deformations are quite different depending on whether one
compactifies the theory on a circle or not.

\section{Off-shell embedding of the fivebrane}\label{sec4}

We have seen that in order to go off-shell we need to embed the
M5-brane non-holomorphically in ${\mathbb R}^5 \times S^1$. The
coordinate $w$ remains a holomorphic function of $v$ and is
parametrized by the $n$ glueball vevs $S_i$, or equivalently by
the coefficients of $f_{n-1}(v)$. The coordinate $t$ on the other
hand becomes non-holomorphic, $t=t(v,\bar v)$, so that the periods
\bea\label{Tperiods}
\oint_{A_i}T&=&2\pi iN_i\nn
\oint_{B_i}T&=&2\pi i\t_i
\eea
Our goal in this section is to study in some more detail the form
of $T$ (or equivalently $t$) implied by (\ref{Tperiods}).

\subsection{General procedure}

The general idea to solve for (\ref{Tperiods}) is to write $T$ as
a linear combinations of closed one-forms. The number of one-forms
compatible with the asymptotic behavior of $T$ is $2n$, so that
the $2n$ equations in (\ref{Tperiods}) indeed completely determine
$T$. There are two different convenient bases of one-forms that we
will use. The first one, which we will call the orthonormal
basis, is useful for many purposes since their $A$-periods are
canonically normalized \cite{farkaskra}. The second one, which we
will call a basis of monomials, is a more natural basis if we are
interested in explicitly integrating $T$ and extracting the
embedding coordinate $t$.

\subsubsection{Orthonormal basis}

As told, $T(v,\bar v)=\frac{\sm{d}t}{ t}$ is off-shell a generic
closed one-form with fixed periods (\ref{Tperiods}). Expanding
upon a basis of holomorphic one-forms, we can write it as
\be\label{expansionT}
T(v)=a\,\tau_{P,\ti P} +b\,\bar\tau_{P,\ti P} +\sum_{i=1}^{n-1}(h_i\xi_i
+l_i\bar\xi_i)
\ee
where $\xi_i$ is a basis of holomorphic differentials
$\{\xi_1,\ldots,\xi_{n-1}\}$ normalized such that
$\oint_{A_i}\xi_j=\d_{ij}$, and which vanish at infinity. In
addition, the A-periods of $\tau_{P,\ti P}$ are chosen to vanish,
and its residue at $P,\ti P$ are $1$ and $-1$, respectively. For a
particular definition of the periods, see appendix
\ref{perintegrals}.

The differentials $\xi_i$ are all of the form $\frac{u_i(v)}{y}
\dd v$ for some polynomials $u_i(v)$ of order at most $n-2$. The
differential $\tau_{P,\ti P}$ is of the form $\frac{u_{n-1}(v)}{y}
\dd v$ with $u_{n-1}(v)$ a polynomial of order $n-1$ with leading
coefficient one. In the monomial basis we will use the
differential forms $\frac{v^i}{y} \dd v$ instead.

Inserting (\ref{expansionT}) in (\ref{Tperiods}) leads to a set of
equations for $a,b,h_i,l_i$. In this computation, one has to bear
in mind that the integral over the non-compact period $B_n$ is
defined as before, by conveniently regularizing the period with
$\L_0$. We find:
\bea\label{conditions}
N &=&a-b \nn
2\pi iN_i &=&h_i-l_i \nn 2\pi i\t_i
&=&\sum_{j=1}^{n-1}(h_i\Pi_{ij} -l_i\bar\Pi_{ij}) +2 \pi i
a\int_{\ti \L_0}^{\L_0}\xi_i - 2 \pi i b\int_{\ti
\L_0}^{\L_0}\bar\xi_i \nn 2\pi i\tau &=&a\int_{\ti
\L_0}^{\L_0}\tau_{P,\ti P} +b\int_{\ti\L_0}^{\L_0}\bar \tau_{P,\ti
P} +\sum_{i=1}^{n-1}\left(h_i\int_{\ti \L_0}^{\L_0}\xi_i
+l_i\int_{\ti \L_0}^{\L_0}\bar\xi_i\right)
\eea
where the last integral is the integral over the non-compact period, and we
used that (see e.g. \cite{csw2})
\be
\frac{1}{2\pi i} \oint_{B_i}\tau_{P,\ti P}=\int_{\ti P}^P\xi_i~.
\ee
The above integrals are not easily computed in general, however,
the leading $\L_0$-dependence is easy to extract. The integrals
$\int_{\ti \L_0}^{\L_0}\xi_i$ and $\int_{\ti
\L_0}^{\L_0}\bar\xi_i$ only give finite contributions as 
$\L_0\rightarrow \infty$. The large $\L_0$ behavior of 
$\int_{\ti \L_0}^{\L_0}\tau_{P,\ti P}$ is
\be \label{ja2}
\int_{\ti \L_0}^{\L_0}\tau_{P,\ti P} = \frac{2}{n+1} \log \left(
\frac{-4 G_n^2(\L_0)}{f_{n-1}(\L_0)}\right) + {\cal
O}\left(\frac{\log\L_0}{\L_0} \right)
\ee
and therefore only involves the leading coefficients of the
polynomials $G_n(v)$ and $f_{n-1}(v)$ that appear in the defining
equation $y^2=G_n^2 + f_{n-1}$ of the Riemann
surface\footnote{This assumes that the leading coefficient of
$f_{n-1}$ is nonvanishing. If it vanishes, and the leading
behavior of $f_{n-1}$ is $v^m$, then the same expression remains
valid except that the prefactor should be changed to $2/(2n-m)$.}.
The only $\L_0$-divergence in (\ref{conditions}) is hence in the
last equation
\be \label{ja1}
2\pi i \tau = a \log \Lambda_0^2 + b \log \bar{\Lambda}_0^2 + {\rm
const} + {\cal O}(\log\L_0/\Lambda_0) .
\ee
On-shell, this equation is directly related to the running of the
bare Yang-Mills coupling as a function of the field theory UV
cutoff: $b=0$ and $a=N$ is proportional to the one-loop beta
function of the field theory. The on-shell value of (\ref{ja2})
can easily be computed using $G_n=P_N$ and
$f_{n-1}=-\Lambda_{N=2}^{2N}$, which yields $2\pi i \tau=\log ( -4
\L_0^{2N}/\L^{2N}_{N=2})$. However, in order to obtain the usual ${\cal 
N}=1$
answers, we will make a slightly different choice, namely
\be \label{ja3}
2\pi i \tau = N\log \left(\frac{-2\L_0^2}{\L_{N=2}^2} \right)
\ee
which would correspond to an embedding of the form $t^2-P_N(v) t +
(-\Lambda_{N=2}^2/2)^N$. This change is just a change of
normalization, and from here on we will use the identification
(\ref{ja3}), which also holds off-shell. If we combine (\ref{ja1})
with (\ref{ja3}) and with the equation $a-b=N$ of
(\ref{conditions}), we obtain
\be \label{ja4}
-N \log \L_{N=2}^2 = b \log (\L_0^2 \bar{\L}_0^2 ) + {\rm const} .
\ee
This shows that $b\rightarrow 0$ as $\L_0\rightarrow\infty$. In
other words, the Riemann surface has a non-holomorphic ``tail'',
which is relevant at the scale of the cutoff $\L_0$. As we send
the cutoff to infinity, the non-holomorphic tail is sent to
infinity as well, but it does yield a finite contribution to
physical quantities for all values of the cutoff. Thus we should first
compute the superpotential and then send the cutoff to infinity, and not
the other way around. Although $b$ vanishes when the cutoff is taken to
infinity, the Riemann surface will in general remain non-homolomorphic in 
the limit: as one can see from (\ref{conditions}), the $l_i$ will in
general remain non-zero.

It is interesting to see how the remaining non-holomorphic deformations, 
parametrized by $l_i$, disappear on-shell. We have already found that 
on-shell $b=0$. Now, on-shell we also have
\be
\tau_i=\sum_{j=1}^{n-1}N_j\Pi_{ij} +N\int_{\ti \L_0}^{\L_0}\xi_i 
=\sum_{j=1}^nN_j\int_{\hat B_j}\xi_i
\ee
where $\hat B_j$ are the non-compact periods, $\hat B_i=B_i+B_n$ for 
$i=1,\ldots,n-1$, and we used the manipulations in \cite{cdsw}. Here we 
regard $S_i$ as the on-shell expectation value of the gluino condensate. 
Combining this with equation \eq{conditions}, we get
\be
\sum_{j=1}^{n-1}l_j\,{\mbox{Im}}\,\Pi_{ij}=0~.
\ee
Notice that by construction ${\mbox{Im}}\,\Pi_{ij}$ is positive definite. So
it has no zero eigenvalues unless $l_i=0$, which is precisely the condition
that all remaining non-holomorphic deformations vanish.

\subsubsection{Monomial basis}\label{sec4.1.2}

The above expansion of the one-form $\dd t/t$ in terms of
canonically normalized one-forms on the Riemann surface allows us to
write its periods in terms of the period matrix of the Riemann surface.
However, if we want to write down an explicit expression for $t$, it
is easier to expand it in the monomial basis
\be
\label{monom}
\zeta_i = \frac{v^{i-1}}{y} \dd v, \qquad 0\leq i \leq n-1.
\ee
The differentials $\tau_{P,\tilde{P}}$ and $\xi_i$ can be written
as linear combinations of the $\zeta_i$, and
the non-normalizable one-form $\zeta_{n-1}$ appears only in
$\tau_{P,\tilde{P}}$ but not in the normalizable one-forms $\xi_i$.
Thus, the one-form $\dd t/t$ is of the form
\be\label{offshellT}
\frac{\dd t}{t}=
T(v,\bar v)=\frac{a_{n-1}(v)}{ y(v)}\,\dd v+\frac{b_{n-1}(\bar
v)}{ \bar y(\bar v)}\,\dd\bar v
\ee
where $a_{n-1}(v)$ and $b_{n-1}(\bar v)$ are polynomials in $v$
that are determined by the conditions on the periods. We will
compute them in some explicit examples below.

It is straightforward to formally integrate $t$ from the expression for
$T(v,\bar v)$. We get:
\be
t(v,\bar v)=\a(v)\b(\bar v)
\ee
where $\a$ and $\b$ satisfy
\bea\label{td}
\frac{\a'(v)}{\a(v)}&=&\frac{a_{n-1}(v)}{ y(v)}\nn \frac{\b'(\bar
v)}{\b(\bar v)}&=&\frac{b_{n-1}(\bar{v})}{\bar y(\bar v)}
\eea
and are formally given by
\bea\label{alphabeta}
\a(v)&=&\exp\int\frac{a_{n-1}(v)}{ y(v)}\,\dd v\nn \b(\bar
v)&=&\exp\int\frac{b_{n-1}(\bar v)}{\bar y(\bar v)}\,\dd\bar v~.
\eea
There is however an important restriction on $\a$ and $\b$, which comes from
the requirement that $t$ must be a well-defined function of $v$ and $\bar
v$.
This means that the product $\a(v)\b(\bar v)$ must be a well-defined 
function
on the Riemann surface. This
is a non-trivial requirement, as the formal integrals \eq{alphabeta} are not
necessarily smooth single-valued functions for all values of $v$, and in
general they will not be. However, the products must be such that $t$ itself
is well-defined as an embedding coordinate. Of course, this is equivalent
to imposing that $T$ has integer periods.

\subsection{Example}

To illustrate the above formalism, we will now apply it to the case where
the gauge group is classically unbroken, so that it confines completely in 
the
quantum theory. In this case the Riemann surface has a single cut and
two punctures.

The Riemann surface is given by the equation
\be
y=\sqrt{(v-c)^2-\m} ~,
\ee
The differential $\dd t/t$ is therefore of the form
\be
\frac{\dd t}{t} = \frac{a}{y}\, \dd v + \frac{b}{\bar{y}}\, \dd\bar{v}
\ee
which can be integrated to give the explicit expression for $t(v,\bar{v})$
\be
t(v,\bar v)=(v-c+y)^a(\bar v-\bar c+\bar y)^b~.
\ee
We can compute the A-period and the regularized B-period of $\dd t/t$, and
this leads (see also (\ref{conditions})) for the A-period to
\be \label{cnd1}
\frac{1}{2\pi i} \oint_A\frac{\dd t}{ t}=a-b=N,
\ee
and for the non-compact B-period we get
\be \label{cnd2}
2\pi i \tau = \int_B\frac{\dd t}{ t}=\log\left(\frac{\L_0-c+y(\L_0)
}{\L_0-c-y(\L_0)}\right)^a \left(\frac{\bar\L_0-\bar c+\bar
y(\bar\L_0)}{\bar\L_0-\bar c -\bar y(\bar\L_0)}\right)^b~.
\ee
For large $\Lambda_0$, this becomes
\be
2 \pi i \tau = \log \left( \frac{4\Lambda_0^2}{\mu} \right)^a
\left( \frac{4\bar{\L}_0^2}{\bar{\mu}} \right)^b  +
{\cal O}(\log\L_0/\L_0)
\ee

For simplicity we will now also assume that the superpotential is
quadratic,
\be W(\Phi) = \frac{1}{2} g_2 (\Phi-c)^2,
\ee
so that the differential $w\dd v$ becomes
\be
w \dd v = g_2 (v-c+y) \dd v.
\ee
The periods are
\be \label{cnd3}
S = \frac{1}{2\pi i} \oint_A w \dd v = -\frac{g_2 \mu}{2}
\ee
and
\bea
\frac{\partial{\cal F}}{\partial S} &  = &
\int_B w\dd v= \frac{g_2}{2\pi i} \left[(\Lambda_0-c)^2 -
\frac{\mu}{2} - \frac{\mu}{2} \log \left(\frac{4\L_0^2}{\mu}\right)\right]+
{\cal O}(\log\L_0/\L_0) \nonumber \\
& = &
W(\L_0) + S - S \log \left( \frac{S}{-2g_2\L_0^2} \right).
\eea
The superpotential (\ref{superpot}) is equal to
\be
W=NS- NS \log \left(
\frac{S}{-2g_2\L_0^2} \right) - 2\pi i \tau S
\ee
where we dropped a constant term proportional to $W(\Lambda_0)$.
By comparing this to the on-shell answer, we see that with our
conventions the relation between $\tau$ and $\Lambda_0$ should be
\be \label{fixtau}
2\pi i \tau = N\log \left( \frac{-2g_2\Lambda_0^2}{\Lambda^3_{N=1}}
\right) =
N\log \left( \frac{-2\Lambda_0^2}{\Lambda^2_{N=2}} \right)~,
\ee
where we used the scale matching $\Lambda_{N=1}^3 = g_2
\Lambda_{N =2}^2$ between the high-energy ${\cal N}=2$ scale
$\Lambda_{N=2}$ and the low-energy ${\cal N}=1$ scale $\Lambda_{N=1}$.
This is exactly the reason why we chose the normalization of
$\tau$ as in (\ref{ja3}). Using (\ref{fixtau}), the superpotential
reduces to the standard Veneziano-Yankielowicz result
\be
W=R  NS - RS \log \left( \frac{S^N}{\Lambda_{N=1}^{3N}}
\right)~,
\ee
and one recovers all familiar results for the ${\cal N}=1$ theory.

Combining (\ref{cnd1}), (\ref{cnd2}), (\ref{cnd3}) and (\ref{fixtau}), we
obtain the following equation for $b$
\be
\label{beq}
b \log \left( \frac{ 4 \L_0^2 \bar{\L}_0^2 g_2^2}{S\bar{S}} \right) =
\log \left( \frac{S^N}{\L_{N=1}^{3N}} \right).
\ee
This result tell us that on-shell $b=0$, since the right hand side of this
equation is then identically zero. However, off-shell $b$ depends 
non-trivially
on $S,\bar{S}$, but also on $\Lambda_0$. As we take the cutoff to infinity,
$b$ becomes smaller and smaller, but this is an example where computing the
superpotential and taking the cutoff to infinity are not two commuting 
operations.
One should first regulate the problem, then compute everything, and finally
take the cutoff to infinity. There is therefore no simple non-holomorphic
surface that gives the right answer without the need to introduce a cutoff.
Finite non-holomorphic surfaces describe situations where the glueball
superfields have expectation values of the order
$|S|\sim |\Lambda_0|^{\eta}$ for some finite value of $\eta$.

The case where the superpotential is arbitrary but the Riemann
surface still has only one cut can also be treated along similar
lines, see e.g. \cite{civ} for more details about this situation.

In appendix~\ref{moreexamples} we illustrate the case of two cuts and comment
on the general case.

\subsection{Alternative variational problem}

So far we have looked at the variational problem where deformations of the 
brane leave the asymptotics of the brane fixed. That is,
we considered normalizable deformations, which are the ones of direct
physical interest\footnote{As remarked, because the Riemann surface was 
non-compact we actually had to include a non-holomorphic deformation with a
logarithmic divergence.}. In principle, however, one can consider holomorphic 
but non-normalizable 
deformations and see what they correspond to in the gauge theory.

This is now straightforward. In this case $t$ is holomorphic, and $w$ 
contains non-normalizable deformations which correspond to changing the 
coefficients of the tree 
level superpotential. The conditions \eq{jj1} are then reformulated as follows:
\bea\label{asymptdefo}
\hat P_N^2-\L^{2N}&=&\hat S_{N-n}^2(\hat G_n+\hat f_{n-1})\nn
\hat W{}_m'{}^2 +\hat{\ti f}_{m-1}&=&\hat H_{m-n}^2(\hat G_n^2+\hat 
f_{n-1})~.
\eea
where again the hat denotes polynomials whose coefficients are to be 
determined from the equations. Again, we need at least $n$ free undetermined
coefficients to vary the $S_i$'s. A simple counting of equations shows that
this is possible.

The above describe the only allowed holomorphic deformations
if we want to keep the polynomial behavior of $t$ and $w$ in $v$ at infinity 
unmodified (even though we change the coefficients of $w$), and without 
introducing additional singularities in $t$. In conclusion, varying the 
$S_i$'s will now mean varying the coefficients of $W'_m$.

This is of course consistent with the fact that the glueball superfields can
be obtained from the effective superpotential, 
$\bra S_i\ket\sim\frac{\pa W_{\sm{eff}}}{ \pa \tau_i}$, as the latter contains
the coefficients of $W'_m$. In view of the partial integration that we did
on Witten's original expression for the superpotential, it would be interesting
to re-analyze the superpotential from this alternative perspective. Notice also 
that the
non-normalizable deformations obtained by
changing the tree-level superpotential of the high-energy theory are reminiscent
of the non-normalizable deformations in AdS that correspond to the sources that 
couple to operators in the CFT.

\section{The superpotential and the (2,0) theory}

One would like to see if the superpotential can be obtained
directly in the (2,0) theory living on the worldvolume of the
fivebrane, as was suggested in \cite{wi2}. For M5-brane
configurations with ${\cal N}=2$ supersymmetry, where only one of
the scalars $t$ or $w$ is active, it is well known \cite{bhoo,hlw}
that the equations of motion for the gauge fields on the 5-brane
correctly reproduce the K\"ahler potential of the ${\cal N}=2$ SYM
theory. It is interesting to see in detail whether this relation
generalizes to the ${\cal N}=1$ case. Of course, once we break
supersymmetry we do not expect the K\"ahler potential to agree,
but we should still be able to reproduce the F-terms and in
particular the superpotential.

So in order to see if we can we get the above superpotential
directly in the (2,0) theory, let us briefly recall what our brane
construction corresponds to in terms of the field content on the
brane. The (2,0) tensor multiplet consists of a (1,0) tensor
multiplet and a hypermultiplet. The bosonic field content of the
(1,0) tensor multiplet is an antisymmetric 2-form potential
$B_{MN}$ with self-dual three-form field strength, and a real
scalar $\f$. The hypermultiplet contains four real scalars which
can be combined into two complex scalars $\f^i$, $i=1,2$. Thus, in
all there are five real scalars which describe the transverse
coordinates to the brane, in our set-up $t$, $w$, and $X^9$ which
we set to zero. Thus only two complex scalars are non-trivial, and
these combine together with the fermions in a four-dimensional
hypermultiplet, which we will denote by $\Phi^i$, $i=1,2$,
following the notation used earlier for the scalars. However, now
one should interpret these as two ${\cal N}=1$ superfields that
make up an ${\cal N}=2$ hypermultiplet. By construction, all
fields in the tensor multiplet are trivial and set to constant
values so we will ignore them for the time being.

To see how to get the superpotential, we can follow the recipe in
\cite{ahgw,dvd} of rewriting the 6-dimensional fields in terms of
an infinite family of 4-dimensional ${\cal N}=1$ superfields
labeled by a continuous parameter, which in this case will be the
complex coordinate $v$ which parametrizes the dependence of the
fields on the internal manifold $\S$. Using the expressions in
\cite{ahgw}, the scalar part of the action reads\footnote{We are
using a linearized version of the Dirac-Born-Infeld action here,
keeping only the quadratic terms. See also the next section.}
\be \label{ssp}
S[\f^i]=\int\dd^4x\,\dd^2v\,\dd^2\th\,\Phi^1\bar{\pa}\Phi^2~.
\ee
There is a choice here which complex coordinate we associate with
$\Phi^1$ and which one with $\Phi^2$. On a compact Riemann surface
both choices are equivalent, but here we are dealing with a
noncompact surface and the two choices are different. As we
explained before, below (\ref{pint}), this is determined by the
boundary conditions, and therefore we will associate $\Phi^1$ with
$w$ and $\Phi^2$ with $\log t$. With this choice we now recognize
that (\ref{ssp}) is the same as our previous expressions
(\ref{pint}) and (\ref{wlogt}) for the superpotential. Notice that
in (\ref{ssp}) we can replace $\bar{\partial}\log t \dd\bar{v}$ by
$\dd\log t$, since the holomorphic piece in $\dd\log t$ does not
contribute to the integral.

Clearly, it would be interesting to study this deconstruction in
more detail and to see whether other terms in the effective action
have an equally nice geometric interpretation.




\section{The resolvent and the gauge kinetic terms}
\label{reskin}

In this section we briefly comment on some other physical
quantities, namely the gauge theory resolvent and its K\"ahler
potential, and their relation to the fivebrane theory.

On-shell, the gauge theory resolvent was simply given by $\dd t/t$, see 
equations
(\ref{per1})--(\ref{rss2}). Off-shell, this can no longer be true, since the
gauge theory resolvent is holomorphic whereas $t$ becomes non-holomorphic. 
It is
however easy to see from (\ref{rss1}) that the A-periods of the gauge
theory resolvent will off-shell still be $N_i$. Therefore, the gauge theory 
resolvent
is given by the unique holomorphic one-form whose A-periods are $N_i$. The 
B-periods
of the gauge theory resolvent will no-longer be integer, but that is not 
important
since the gauge theory resolvent need not have a straightforward geometrical
interpretation. To compute the gauge theory resolvent we can either 
construct
directly a suitable one-form, or we can use
\be
{\rm Tr}_{\rm gauge theory} \left( \frac{\dd v}{v-\Phi} \right) =
\sum_i N_i \frac{\partial}{\partial S_i} w \dd v
\ee
which follows from (\ref{rss1}) and (\ref{rss2}), or equivalently we can see
this directly from (\ref{per1}). It would be interesting to understand the
gauge theory resolvent more directly from the fivebrane point of view.

Let us next consider the K\"ahler potential of the low-energy effective
theory on the five-branes, as considered in \cite{bhoo,hlw}. We will mostly
follow the computation of \cite{bhoo}. The K\"ahler potential is not 
protected
by supersymmetry and therefore we don't expect the answer that we find to be
in agreement with the K\"ahler potential in the pure gauge theory. 
Nevertheless,
it may share some qualitative features with the gauge theory answer and it 
is
therefore worth exploring.

Since the off-shell fivebrane configuration is non-holomorphic, the induced 
metric
on it will not be proportional to $\dd v \dd \bar{v}$, but also have
$\dd v \dd v$ and $\dd \bar{v}\dd\bar{v}$ components. In principle, it would 
be
nicer to introduce different coordinates on the Riemann surface in terms of
which the metric is in conformal gauge, but in practice it may be difficult 
to
explicitly construct such coordinates.

To write the kinetic term for the glueball fields, we write the metric in
the internal directions as
\be
G_{pq}\dd X^p \dd X^q = \dd v \dd \bar{v} + \dd w \dd \bar{w} +R^2\dd \log t 
\dd \log \bar{t}.
\ee
We will also write $v^{\alpha}=(v,\bar{v})$, and for the Riemann surface
take a family that varies over ${\mathbb R}^4$, i.e. one where the glueball
fields are non-constant. In other words, we take as our Riemann surface
$X^p \equiv X^p(v,\bar{v},S^i(x^{\mu}),\bar{S}^i(x^{\mu}))$.

We define
\bea
E_{\mu}^p & = & \partial_{\mu} S^i \partial_i X^p + \partial_{\mu} 
S^{\bar{i}}
\partial_{\bar{i}} X^p \nonumber \\
g_{\alpha\beta} & =  &\partial_{\alpha} X^p G_{pq} \partial_{\beta} X^q
\eea
and
\be
H_{pq}= G_{pr} \partial_{\alpha} X^r g^{\alpha\beta} \partial_{\beta} X^s 
G_{sq}.
\ee
The metric $g_{\alpha\beta}$ is the induced metric on the Riemann surface.
The kinetic terms for $S^i,\bar{S}^i$, as obtained from the Born-Infeld 
action,
now read
\be \label{ja9}
S_{\rm kin} = \int d^4 x d^2 v \sqrt{-\det g_{\alpha\beta}}\,
\eta^{\mu\nu} E_{\mu}^p (G_{pq}-H_{pq}) E_{\nu}^q .
\ee
This is a rather complicated expression due to the fact that the fivebrane
configurations are non-holomorphic. A puzzling feature is the appearance
of kinetic terms of the form $\partial_{\mu} S^p \partial^{\mu} S^q$ that 
include
two holomorphic fields $S$ (and similar terms with two antiholomorphic 
glueball fields).
In an ${\cal N}=1$ theory, one would expect only terms of the form
$\partial_{\mu} S^p \partial^{\mu} \bar{S}^q$. Here, the fivebrane 
configuration explicitly
breaks ${\cal N}=1$ supersymmetry, and this is why terms like
$\partial_{\mu} S^p \partial^{\mu} S^q$ are being generated. It is possible 
that this is
an artifact of the fivebrane theory, or that due to some mysterious 
cancellation
these terms disappear after integrating over the Riemann surface. In any 
case, we cannot
derive a K\"ahler potential from (\ref{ja9}). The only way to obtain a 
K\"ahler potential
from (\ref{ja9}) is to ignore all $\bar{v},\bar{S}$ dependence in $\log t$. 
That leads to
a K\"ahler potential
\be \label{ja10}
K \sim \int \dd^2 v \, k \partial_{v} \partial_{\bar{v}} k
\ee
where
\be
k= |v|^2 + |w|^2 + R^2 |\log t|^2
\ee
is the space-time K\"ahler potential, now viewed as a function of 
$v,\bar{v},S,\bar{S}$.
There may be some relation between the field theory K\"ahler potential and 
(\ref{ja10}),
but there is no a priori reason to believe such a relation exists.

\section{Conclusion}

In this paper we have argued that the off-shell deformations of the M5-brane 
that
correspond to turning on non-trivial expectation values for the glueball 
fields
are particular non-holomorphic ones. We have expressed these in terms of the
periods of differentials made out of the embedding coordinates and used this 
to
compute the superpotential of these theories using the definition of the
superpotential  given by Witten in \cite{wi1}. This definition
involves the integral over a three-manifold
that bounds two different Riemann surfaces and can be viewed as computing
a domain wall tension, i.e. the difference between the value of
the superpotential at two different vacua.
By partial integration, it can be written as an integral over a
two-form on $\Sigma$; such a form was also used in e.g. \cite{av,kach,ler}.
This form has many interpretations. It can be viewed as the contribution
of a 6d hypermultiplet to the superpotential as we discussed above,
equivalently one can obtain it from a dimensional reduction of the
type IIB superpotential $\int H\wedge \Omega$. All
these give the same answer for the superpotential.
We also saw that after carefully taking the boundary conditions into
account, we could recover both the DV-version of the superpotential,
as well as a more direct description in terms of the gauge theory
resolvent.

Altogether, we see that the single M5-brane picture provides a setup
where everything has become geometrical, and in particular all
quantum information has become geometrical. The computations bear a close
resemblance to the computations in IIB string theory
with fluxes where one employs the superpotential $\int H\wedge\Omega$.
In fact, the IIB string theory and the M5-brane configuration are T-dual
to each other; under this T-duality, non-trivial IIB geometry and flux
are replaced by a trivial background geometry with a non-trivial brane
configuration.

The approach in \cite{dv1} is somewhat intermediate, in that
part of the quantum dynamics is encoded in the matrix model
that arises. It is not quite clear in what sense this approach, which
uses topological string theory, is dual to or can be embedded directly in 
the
M5-brane setup. The action of the matrix model itself is similar to the
IIA brane configuration, in that it only has classical information, and 
doing
the matrix integral is analogous to the lift of the IIA configuration to 
M-theory.
The precise relation between the M5-brane physics and the topological string
theory is clearly something worth pursuing further.

There are many directions in which the results of this paper can be
generalized. It should be relatively straightforward to generalize these 
results
to other gauge theories for which a IIA brane description is known, such as 
theories
with matter, quiver gauge theories and theories with $SO(N)/SP(N)$ gauge 
groups.
It would certainly also be worthwhile to see to what extent non-holomorphic 
deformations
of other wrapped brane configurations are physically relevant.
We also believe that this work should have some relation to and implications 
for
more general topological string backgrounds such as the the ones considered
in \cite{vertex1,vertex2}, as well as for the approach to ${\cal N}=2$ 
theories
advocated in \cite{nekr}. We hope to return to some of these issues in the
near future.

\section*{Acknowledgments}

We would like to thank Robbert Dijkgraaf and Annamaria Sinkovics
for collaboration and useful discussions during the early stages
of this work. We would also like to thank Mina Aganagic, Tim
Hollowood and Cumrun Vafa for valuable discussions and
suggestions. SdH thanks the Simons Workshop on Mathematics and
Physics, where part of this work was carried out, and the
Institute for Theoretical Physics of the University of Amsterdam
for hospitality.

This material is based upon work supported by the National Science 
Foundation
under Grant No. PHY-0099590. Any opinions, findings, and
conclusions or recommendations expressed in this material are those of the
authors and do not necessarily reflect the views of the National Science
Foundation.

\appendix
\section{Definition of the period integrals}\label{perintegrals}

Integrals along paths going from one branch of the Riemann surface to the
other are a priori not well defined, as one in addition needs to specify the
path. Here we briefly give our definition of these integrals by a definition
of the A- and B-periods that we use in practical computations.

A basis of compact A-cycles is provided by the cycles $A_i$,
$i=1,\ldots, n-1$, encircling a single cut in the counterclockwise direction
without intersecting. In our applications, however, we also need to take
into account the points at infinity, $P$ and $\ti P$, and so we will have to
supplement this with integrals around $P$. Notice that in our applications
the sum over all A-periods, including $A_n$, can be shown to be equal to the
contour around $P$ by a contour deformation. This (infinite) contour is
defined by $\L_0\rightarrow e^{2\pi i}\L_0$, where $\L_0$ is a regulator 
that
we take to infinity at the end.

Concretely, we compute A-periods by taking the contours counterclockwise and
defining the integrals with the appropriate signs, depending on which branch
we are on:
\be
\oint_{A_i}=\int_{\ti c_i^-}^{\ti c_i^+} +\int_{c_i^+}^{c_i^-}~,
\ee
$i=1,\ldots,n$. In most applications we encounter, these integrals add up
to give a factor of 2.

We can choose a set of compact B-cycles $\{B_1,\ldots,B_{n-1}\}$ defined as
the cycles going from one cut to the other in the clockwise direction:
\be
\oint_{B_i}=\int_{c_i^+}^{c_{i+1}^-} +\int_{\ti c_{i+1}^-}^{\ti c_i^+}~,
\ee
$i=1,\ldots,n-1$. Also here we need to supplement this with a non-compact
period, which we take as going from $\ti P$ to $P$ passing through $c_n^+$
and appropriately regularized:
\be
\int_{B_n}=\int_{\ti\L_0}^{\ti c_n^+} +\int_{c_n^+}^{\L_0}~.
\ee
In the case where $n=1$, this will be the only period and we will call it
simply $B$.

%
%
%

\section{The case with more than one cut}
\label{moreexamples}

In this appendix we illustrate the method outlined in section \ref{sec4} for
surfaces with more cuts. We look at the case corresponding to the
symmetry breaking pattern $U(N)\rightarrow U(N_1)\times U(N_2)$ in the field
theory, or $n=2$. So we write
\be
y^2(v)=(v-c_1^-)(v-c_1^+)(v-c_2^-)(v-c_2^+)
\ee
In this case we will need the following two basic integrals in
order to solve for $T$:
\bea\label{ellint}
&&\int\frac{\dd v}{ y(v)}\nn &&\int\frac{v\dd v}{ y(v)}
\eea
These integrals are given in terms of elliptic integrals of the
first and third kind\footnote{As an illustration, it is useful to
first consider the case where the minima $c_i$ are such that we
can use the following parametrization:
$y=\sqrt{(t^2-\m_1)(t^2-\m_2)}$, which is consistent with the fact
that $f_{n-1}(v)$ has two coefficients in this case. Using
rescalings we can rewrite the integrals over the periods in terms
of $\int_0^1\frac{\dd v}{\sqrt{(1-v^2)(1-k^2v^2)}}=K(k)$ and
$\int_1^{1/k}\frac{\dd v}{\sqrt{(1-v^2)(1-k^2v^2)}}=iK'(k)$, where
$K$ and $K'$ are the usual complete elliptic integrals of the
first kind \cite{Lawden}. The second integral in \eq{ellint} can
be explicitly computed in terms of a simple logarithm.}.

For simplicity we will work in the classical limit, where the
distance between the two cuts is much larger than their length,
and we take $c_i^\pm=c_i\pm\m_i$, where $\m$ is a small parameter
related to the coefficients $f_i$. Our approximation will be an
expansion in $\frac{\m_i}{ c_1-c_2}$. In the field theory, since roughly
$\m\sim\L^{2N}_{N=2}$, this corresponds to the classical limit of small
mass scale $\L_{N=2}$.

We can write an integral over the first cut as:
\be
\oint_{A_1}\frac{\dd v}{
y(v)} =\frac{2}{ c_1-c_2}\int_{-\sqrt{\m_1}}^{\sqrt{\m_1}}
\frac{\dd x}{\sqrt{x^2-\m_1}} +{\cal O}\left(\frac{\m_1}{
c_1-c_2}\right)= -\frac{2\pi i}{ c_1-c_2}+{\cal O}
\left(\frac{\m_1}{ c_1-c_2}\right)
\ee
where we took the period in the counterclockwise direction.
Obviously, there is a similar contribution around $A_2$. We also
need the following integral:
\be
\oint_{A_i}\frac{v\dd v}{ y(v)}=c_i\oint_{A_i}\frac{\dd v}{ y(v)}
+{\cal O}\left(\frac{\m_i}{ c_i-c_j}\right)~.
\ee
Writing
\be
T=(a_0+a_1 v)\,\frac{\dd v}{ y(v)} +(b_0+b_1\bar v)\,\frac{\dd\bar
v}{\bar y(\bar v)}
\ee
and taking into account all contributions we get
\be\label{Nis}
N_i=-\frac{a_0+a_1c_i}{ c_i-c_j}+\frac{b_0+b_1\bar c_i}{ \bar
c_i-\bar c_j} +{\cal O}\left(\frac{\m}{ c_i-c_j}\right)
\ee
where obviously $j\not=i$.

The $\L_0$-dependence of the non-compact period $B_2$ can be
computed as before in powers of $\log\L_0/\L_0$. We get:
\be
\int_{\ti\L_0}^{\L_0} T=2a_1\log\L_0 +2b_1\log\bar\L_0+{\cal
O}(\log\L_0/\L_0) =2\pi i\t.
\ee
This, together with \eq{Nis} and the integral over $B_1$, which is
again given by an elliptic integral, determines
$a_0,a_1,b_0$ and $b_1$ uniquely.

Let us make some comments about the general case,
$U(N)\rightarrow\prod_{i=1}^nU(N_i)$. As the genus of the curve goes up, the 
integrals that one needs to
solve get more and more involved. It is however not hard to see
how the general case works. One needs to impose:
\bea\label{off-shellperiods}
\oint_{A_i}\frac{a_{n-1}(v)}{ y(v)}\,\dd v
+\oint_{A_i}\frac{b_{n-1}(\bar v)}{\bar y(\bar v)}\,\dd\bar
v&=&2\pi iN_i\nn \oint_{B_i}\frac{a_{n-1}(v)}{ y(v)}\,\dd
v+\oint_{B_i}\frac{b_{n-1}(\bar v) }{\bar y(\bar v)}\,\dd\bar
v&=&2\pi i\t_i
\eea
both for the normalizable periods and for the non-normalizable
one. The integrals over the normalizable periods are integers, as
follows from the fact that $T$ is on-shell a meromorphic
differential that is written locally as $\dd\log t$ where $t$ is
the embedding coordinate. And even though $T(v,\bar v)$ takes on a
similar form by the relations \eq{td}, it is crucial that $\a(v)$
and $\b(\bar v)$ are not rational functions. Indeed, as we have
seen explicitly, the exponents $a$ and $b$ are not integers, but
they can take any values as functions of $N,\L_0,\m$, and $c$.

It is now useful to see whether the conditions
\eq{off-shellperiods} have solutions for arbitrary values of $n$.
One can easily see that this is the case. They form $2n$
conditions, and we have $2n$ coefficients, $\{a_1,\ldots,a_n\}$
and $\{b_1,\ldots,b_n\}$, to adjust, so the solution, if there is
one, is unique. It is thus crucial that we also include one
off-shell log-normalizable deformation $b_{n-1}$.

On-shell, the $B$-periods are fixed once the complex structure
$f_i$, $i=0,\ldots,n-1$, is fixed. In that case we can set all
$b_{n-1}(\bar v)=0$, and we are left with the $n$ parameters of
$a_{n-1}(v)$, as we showed in section 4.

The general case can be done without much more effort. If one
wants to stay completely general, the expressions become
complicated elliptic integrals. However, one can easily
approximate these integrals in the semi-classical limit we have
been considering before. We will skip details and just quote the result
for the glueball expectation values:
\be\label{Sis}
S_i=H_{m-n}(c_i)\,\frac{f_{n-1}(c_i)}{2G'_n(c_i)}~,
\ee
Using the monodromy argument in \cite{civ} one also finds
\be
\ti
f_{m-1}=2\sum_{i=1}^nS_i=\sum_{i=1}^nH_{m-n}(c_i)\,\frac{f_{n-1}(c_i)
}{ G'_n(c_i)}~.
\ee
One can compute the A-periods of $T$ and the leading $\L_0$-dependence of 
the
$B$-periods of $T$ and $w\dd v$ similarly.

\end{document}